\documentclass[reprint,superscriptaddress]{revtex4-1}
\usepackage{amsmath,amssymb,units,amsfonts}
\usepackage[dvipdfmx]{graphicx}
\usepackage{color}
\usepackage{verbatim}
\usepackage{tikz}
\usepackage{siunitx}
\usepackage{bm}

\begin{document}
\title{Magnetic solitons in a spin-1 Bose-Einstein condensate}

\author{X. Chai}
\thanks{These authors contributed equally to this work.}
\affiliation{School of Physics, Georgia Institute of Technology, 837 State St, Atlanta, Georgia 30332, USA}
\author{D. Lao}
\thanks{These authors contributed equally to this work.}
\affiliation{School of Physics, Georgia Institute of Technology, 837 State St, Atlanta, Georgia 30332, USA}
\author{Kazuya Fujimoto}
\affiliation{Institute for Advanced Research, Nagoya University, Nagoya 464-8601, Japan}
\affiliation{Department of Applied Physics, Nagoya University, Nagoya 464-8603, Japan}
\author{Ryusuke Hamazaki}
\affiliation{Department of Physics, University of Tokyo, 7-3-1 Hongo, Bunkyo-ku, Tokyo 113-0033, Japan}
\author{Masahito Ueda}
\affiliation{Department of Physics, University of Tokyo, 7-3-1 Hongo, Bunkyo-ku, Tokyo 113-0033, Japan}
\affiliation{Institute for Physics of Intelligence, University of Tokyo, 7-3-1 Hongo, Bunkyo-ku, Tokyo 113-0033, Japan}
\affiliation{RIKEN Center for Emergent Matter Science (CEMS), Wako, Saitama 351-0198, Japan}
\author{C. Raman}
\email {Corresponding author:  craman@gatech.edu}
\affiliation{School of Physics, Georgia Institute of Technology, 837 State St, Atlanta, Georgia 30332, USA}

%
%
%
%

\begin{abstract}
	\noindent {\bf Abstract.} Vector solitons are a type of solitary, or non-spreading wavepacket occurring in a nonlinear medium comprised of multiple components.  As such, a variety of synthetic systems can be constructed to explore their properties, from nonlinear optics to ultracold atoms, and even in human-scale metamaterials.  In quantum systems such as photons or Bose-Einstein condensates (BECs), such vector nonlinearities offer a window into complex many-body dynamics, and offer possibilities for quantum communication and information processing.  BECs have a rich panoply of internal hyperfine levels, or spin components, which make them a unique platform for exploring these solitary waves.  However, existing experimental work has focused largely on binary systems confined to the Manakov limit of the nonlinear equations governing the soliton behavior, where quantum magnetism plays no role.  Here we observe, using a ``magnetic shadowing'' technique, a new type of soliton in a spinor BEC, one that exists only when the underlying interactions are antiferromagnetic, and which is deeply embedded within a full spin-1 quantum system.  Our approach opens up a vista for future studies of ``solitonic matter'' whereby multiple solitons interact with one another at deterministic locations, and eventually to the realization of quantum correlated states of solitons, a longstanding and unrealized goal.

\end{abstract}
\maketitle

\section{Introduction}

Ultracold atoms have opened a new arena for the exploration of nonlinear behavior with unique experimental tools available.  A prime example of this has been the fruitful study of soliton nonlinearities  \cite{Burger1999,Denschlag2000,Dutton2001,Haljan2001,stre02,Khaykovich2002,Cornish2006,Marcovitch2008,Landa2014}.  Solitons are ubiquitous in the natural world, from the realm of shallow water waves \cite{Gardner1967} to biological systems \cite{Dauxois2006} and even extending into early universe cosmology \cite{Ansoldi2007}.  In synthetic nonlinear systems, vector solitons have been observed in human-scale metamaterials \cite{Deng2019}, and have been proposed as a means to control dispersion in fiber optics, with practical applications for optical communication \cite{Trillo1988,Christodoulides1988}.  In quantum systems such as photons or Bose-Einstein condensates (BECs), such vector nonlinearities offer a window into complex many-body dynamics, and offer possibilities for quantum communication and information processing \cite{Steiglitz2017}.  The multiple flavors available in Bose gases due to the rich variety of internal hyperfine spin components has enabled a number of  theoretical and experimental works on vector solitons \cite{Busch2001,Becker2008,Weller2008,Frantzeskakis2010,Hamner2011,Yan2012,Danaila2016}.  Till now, however, such works have largely explored the Manakov limit of the nonlinear equations where all species are essentially treated equally, and the differences between them are of no consequence.  Little seems to be known about the physics outside of this regime, i.e. the connection between binary solitons and higher spin objects, including $F=1$ spinors, where the underlying magnetic interactions between species play an important role \cite{Stamper-Kurn2013}.  Very recent theoretical work has explored polarization waves outside of the Manakov limit \cite{Kamchatnov2014}, and found  an exact solution under the assumption of a uniform total density \cite{Qu2016}.  It has also explored the connection between solitons and thermalization of nonequilibrium Bose gases \cite{Fujimoto2019}, where spin-spin interactions play an important role.  However, only one such experiment has been reported \cite{Bersano2018}, and a comprehensive description of nonlinear phenomena including these magnetic interactions has yet to emerge.

This work takes an important step forward by providing the first experimental evidence of the magnetic soliton predicted by \cite{Qu2016} in a quasi-one dimensional sodium spinor Bose-Einstein condensate.  We use a magnetic phase imprinting method  to experimentally create solitons in a two component $F=1,m_F = \pm 1$ hyperfine mixture in the antiferromagnetic spinor phase \cite{Stamper-Kurn2013}. To our knowledge, this method has only been explored numerically \cite{Xiong2010}.  In contrast to previous work, the solitons we have created depend crucially on antiferromagnetic interactions between the spins.  A powerful tool at our disposal is the availability of local, {\em in-situ} spin measurements that access the full three-component hyperfine manifold in order to probe the phase profile of the solitons in a manner not typically possible with binary mixtures.

\section{Results}

\begin{figure*} [htbp]
\includegraphics[width =1.5 \columnwidth]{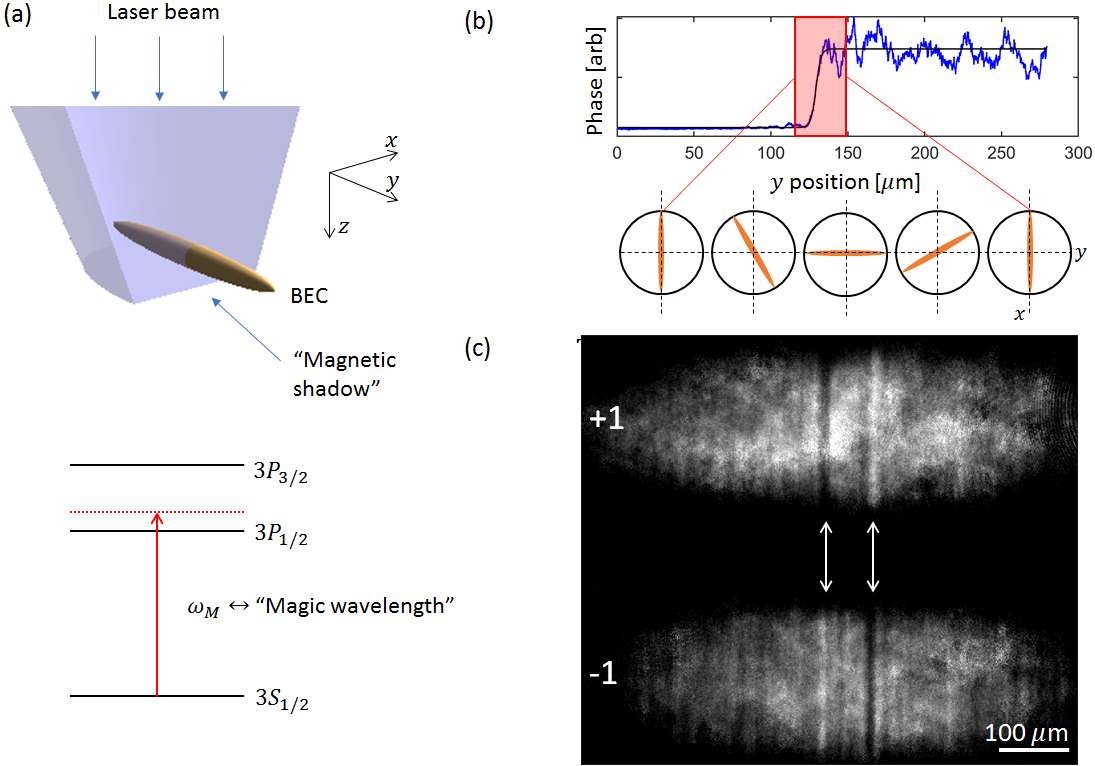}
\caption{(Color Online).  Using a magnetic shadow to create magnetic solitons. (a) A sodium BEC prepared in an equal superposition of $m_F = \pm 1$ hyperfine states is illuminated by a pulse of far-off resonance, circularly polarized light that creates an effective magnetic field due to vector light shifts.  The use of a magic wavelength eliminates scalar shifts of the $3$S$_{1/2}$ level due to destructive interference between contributions from the $3$P$_{1/2}$ and $3$P$_{3/2}$ levels.  A knife edge placed in the beam is imaged onto the atom cloud, resulting in a ``magnetic'' shadow whose edge width (10\% to 90\%) is 8 $\mu$m. (b) Across the 8 $\mu$m transition region,  the effective magnetic field gradient causes differential Larmor precession that, after 120 $\mu$s, results in a $2\pi$ phase winding and a magnetic instability.  Shown schematically is the gradient-induced twisting of the nematic director for these two spin states, where $\alpha$ is the phase difference between $m_F = \pm 1$ components.  (c) Magnetic soliton formation at $t = 20$ ms after application of the pulse. Due to the global conservation of spin, a positive and negative pair of magnetic solitons are created.  Positively magnetized solitons are seen as a density hump (dip) in the $+1 (-1)$ clouds on the right side of the cloud, and a corresponding negatively magnetized soliton on the left side of the cloud.  Images were taken after 9 ms time-of-flight with Stern-Gerlach separation.}
\label{fig:experimental_setup}
\end{figure*}

Here we describe the background to our work and present our main experimental results.  Bose-Einstein condensates in two internal levels $1$ and $2$ exhibit a variety of nonlinear behavior through the coupled Gross-Pitaevskii equations:
\begin{eqnarray}
i \hbar \frac{\partial}{\partial t}\Psi_1 = \left ( -\frac{\hbar^2}{2M} \nabla^2 + V + g_{11}|\Psi_1|^2 + g_{12}|\Psi_{2}|^2 \right ) \Psi_1, \nonumber\\  \\
i \hbar \frac{\partial}{\partial t}\Psi_2 = \left ( -\frac{\hbar^2}{2M} \nabla^2 + V + g_{22}|\Psi_2|^2 + g_{12}|\Psi_{1}|^2 \right ) \Psi_2, \nonumber \\
\label{eq:GP}
\end{eqnarray}
where  $\Psi_{i}$ is the wavefunction of level $i=1,2$ and $V$ is an external trapping potential. 
The strength of a repulsive interaction between atoms in levels $i$ and $j$ is denoted by $g_{ij}$. For bosonic alkali atoms, these two levels are usually two hyperfine states and the interaction strengths are nearly identical. 
Then if $g_{ii}$ is set equal to $g_{ij}$ the above equations reduce to the Manakov limit, where the nonlinear terms can be replaced by a single constant $g \times $ the total density $n = |\Psi_1|^2 + |\Psi_2|^2$.  In this limit a variety of nonlinear solitary waves exist, including dark-bright solitons, where a small domain of one species is trapped within a density minimum of the other \cite{Busch2001,Becker2008,Hamner2011,Yan2012}.  

In this work, we explore nonlinear waves outside of this Manakov regime.  We exploit the small, yet finite and positive difference in scattering length that exists for the mixture of $m_F = +1$ and $m_F = -1$ levels of the $F = 1$ hyperfine ground state (states $1$ and $2$, respectively).  These two spin states are created at the beginning of the experiment by a fast, 160 $\mu$s RF pulse applied to a BEC in the $m_F = 0$ state (see Methods section).  These being equal and opposite spin projections of the vector spin-1 operator $\hat{{\bf F}}$, symmetry causes the intraspecies interactions $g_{11} = g_{22} \equiv g$ to be equal at low magnetic fields \cite{Stamper-Kurn2013}.  The small difference $\delta g \equiv g-g_{12} \approx +0.07 g$ as determined by Feshbach spectroscopy \cite{Knoop2011}.  Since $g_{12}< \sqrt{g_{11} g_{22}}$ a two component mixture is miscible in its ground state, and will not phase separate.  For zero total magnetization $m_z= 0$ and negative quadratic Zeeman shift \cite{Stamper-Kurn2013}, this ground state is an easy-plane nematic, consisting of an equal superposition of states 1 and 2.  It has a director $\vec{d}=(\cos{\phi},\sin{\phi},0)$ that lies in the $x-y$ plane, making an angle $\phi$ with the $x$-axis, as seen in Figure \ref{fig:experimental_setup}(b).  In terms of the relative phase $\alpha$ between the two spin components that we control in the laboratory, there is a simple relation $\phi = (\pi-\alpha)/2$.

Although the order parameter is principally nematic, i.e.\ without any magnetization, nonetheless there can be magnetic excitations in which the density difference $n_1 - n_2$ is nonzero.  One such excitation is a magnetic soliton.
As predicted by Qu et al.\ \cite{Qu2016}, it consists of a density hump of one species atop a density dip in the other, so that in one dimension $y$, the density profile of each species is given by
\begin{equation}
n_{1,2} = \frac{n_0}{2}  \left [ 1 \pm \sqrt{1-U^2} {\rm sech}\left (\frac{\sqrt{1-U^2}}{\xi_{sp}} (y - V t) \right ) \right ] 
\label{eq:soliton}
\end{equation}
where $n_0$ is the background total density and $U = V/c_s$ the soliton velocity normalized to that of spin waves, $c_s = \sqrt{n_0 \delta g/(2 m)}$.  The width of the soliton depends upon the spin healing length $\xi_{sp} = \hbar/\sqrt{2 m n_0 \delta g}$.  For our cigar-shaped BEC with aspect ratio $\approx 70$, we average the density profile over the two transverse dimensions to obtain $\bar{c}_s = c_s/\sqrt{2} = 1.47\pm0.17$ mm/s and $\bar{\xi}_{sp} = \sqrt{2}\xi_{sp} =  0.92\pm0.10~\mu\mathrm{m}$, respectively, using standard time-of-flight expansion and the value of $\delta g/g = 0.07$ given in Reference \cite{Knoop2011}.  Hereafter  $c_s$ and $\xi_{sp}$ refer exclusively to the radially averaged parameters without notating the bar.

The above soliton appears superficially similar to dark-bright Manakov vector solitons in which one component is trapped within another \cite{Busch2001}.  However, there are critical differences between the two.  For one, we create solitons in a mixture of two otherwise miscible spin components, unlike the Manakov soliton of Reference \cite{Busch2001}.  Moreover, its stability does not derive from {\em trapping} one species within another, but from an antiferromagnetic spin-exchange interaction as can be seen from Eq. (3) (note that $\delta g$ must be positive there).


\begin{figure} [htbp]
\includegraphics[width =\columnwidth]{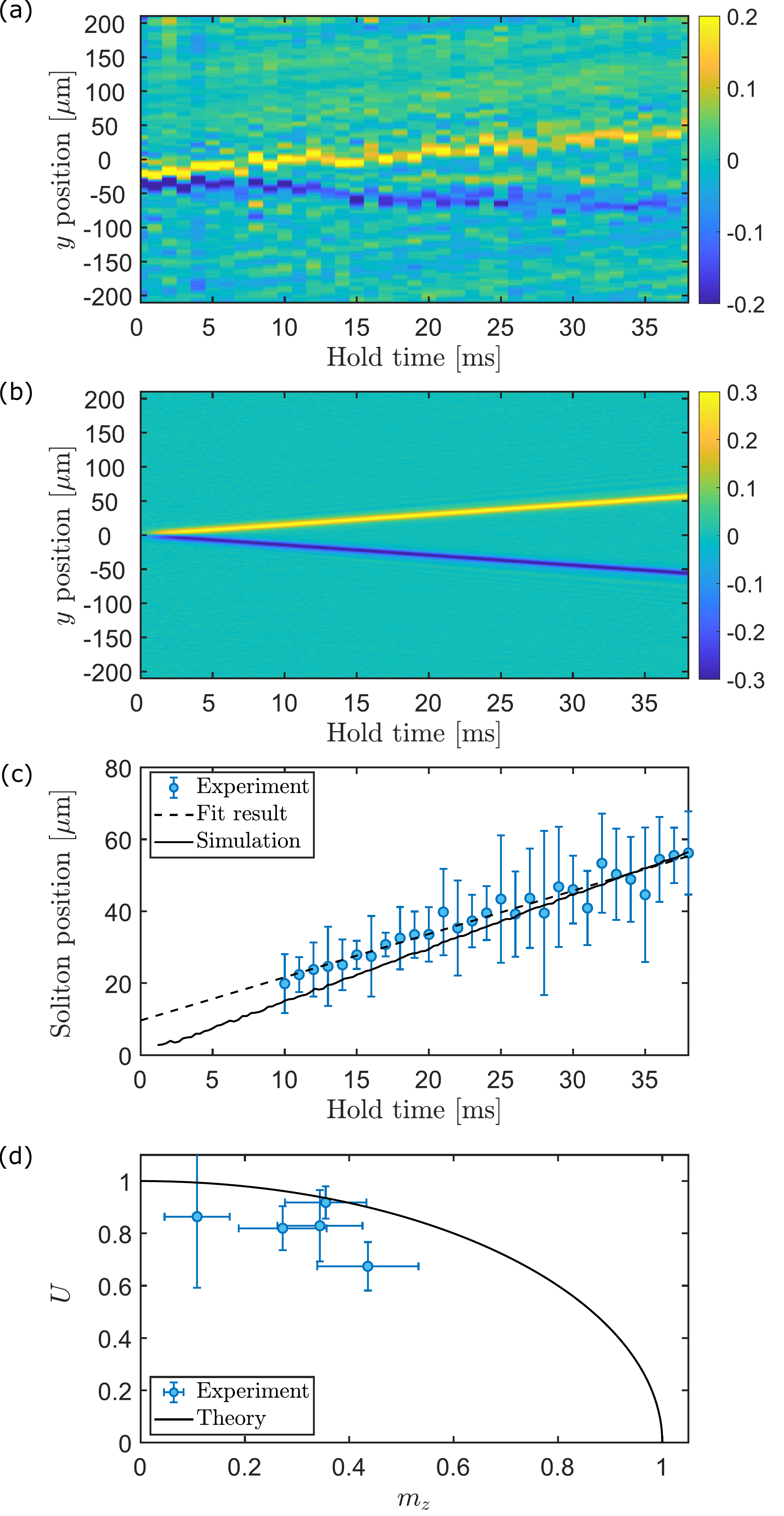}
\caption{(Color Online).  Magnetic soliton formation and propagation.  (a) Spacetime plot of magnetization profiles $m_z(y,t)$ for different hold times after phase imprinting.  
The initial phase step is rapidly converted into a positive and negative magnetization pulse that begin to separate by $\approx 9$ ms. Two slowly moving, unipolar magnetic structures (the magnetic solitons) with velocity $v_S = \pm 1.20$ mm/s were observed. (b) Numerical simulation by solving the 1D GP equation for the same parameters used in the experiment. (c) Soliton position versus time for both experiment (circles) and theory (solid black line).  Dashed line is a straight line fit to the experimental data. (d) Comparison of soliton velocity $U=V/c_s$ and peak magnetization $m_z$, where $V$ is the soliton's velocity and $c_s$ is the spin sound velocity.  Data were obtained by varying the imprinted phase, as discussed in the text.  Experimental error bars are $\pm 2\sigma$ uncertainty in the measured velocity.  Solid line shows the universal curve $U = \sqrt{1-m_z^2}$.}
\label{fig:main_result}
\end{figure}

The relative phase between the two species undergoes a $\pi$-phase jump across the soliton. In contrast to the scalar dark soliton, a richer dynamics ensues through the fractional magnetization $m_z = (n_1-n_2)/(n_1+n_2)$.  $m_z$ can either be positive or negative, independent of the soliton's velocity, which is determined by the sign of the phase jump ($\pm \pi$).  However, due to overall spin conservation, the sign of the soliton's magnetization and number of solitons will depend on the global spin imbalance that has been prepared in the system. The soliton's phase jump can also be quantified through the nematicity $N_{ij} \equiv \frac{1}{2}\langle \hat{F}_i \hat{F}_j + \hat{F}_j \hat{F}_i\rangle$, which is a tensor constructed from the local axis of the nematic director, i.e. $N_{ij} = \delta_{ij}-d_i d_j$.  For the above soliton, it can be shown that $N_{xx}$ experiences a steep domain wall at $y = Vt$ \cite{Son2002}.  We have performed numerical simulations in one dimension that confirm that if $\delta g \rightarrow 0$ the above soliton solution will no longer be stable.

The use of magnetically sensitive levels to create and characterize solitons presents unique experimental challenges.  At typical laboratory fields of 70 mG, there is a factor of 1000 in the ratio of the soliton dynamical timescale ($\sim 20$ ms) to that of Larmor precession ($\sim 20$ $\mu$s).  Controlling the absolute phase difference would require bias stability at the $0.1$\% level, or $\approx 70 \mu$Gauss, below the field instability of our setup.  To overcome this limitation, we use the effective magnetic field generated by a far detuned laser beam that is partially shadowed by a sharp knife edge to imprint a phase gradient across the atom cloud, which does not require control over the absolute value of the phase $\alpha$, only its variation in space.

We illustrate this ``magnetic shadowing'' technique in Figure \ref{fig:experimental_setup}(a).  It utilizes the vector light shift of a circularly polarized laser beam at a ``magic wavelength'' where the scalar light shifts from the $3$P$_{1/2}$ and $3$P$_{3/2}$ levels cancel one another \cite{Arora2007}. Thus we only changed the relative phase between the the two states, and not their phase sum.  This means the phase imprinting does not couple to the overall density, but only to the magnetization, and thus no dark solitons were created as in \cite{Burger1999,Denschlag2000,Becker2008,Frantzeskakis2010}.  The magnetic shadow induces differential Larmor precession rates across a small region of space in the vicinity of the image of the knife edge.  We use a 120 $\mu$s light pulse duration to engineer a $2 \pi$ relative phase step (see Methods section).  This phase step is unstable due to the opposite momentum imparted to $m_F = \pm 1$ atoms.  It decays into a pair of magnetic solitons with opposite value of $m_z$.  The global conservation of magnetization therefore plays an important role in the dynamics of phase engineering--without processes that change the global spin imbalance $n_1-n_2$, the solitons must be produced in pairs of equal and opposite magnetization.  Figure  \ref{fig:experimental_setup}(c) shows time-of-flight Stern-Gerlach images of the two spin states 20 ms after the phase imprinting pulse. It shows the equal and opposite density regions in the two spin states associated with the two solitons.  Similar work on dark-bright solitons has used the counterflow instability to spontaneously form solitons where one species is trapped within another \cite{Hamner2011}.  Our approach, by contrast, is deterministic, since we can place the solitons at the location of the knife edge at an exact instant of time.  This opens up new possibilities for magnetic soliton engineering and studies of their dynamics.  We note that the magic wavelength has been used to excite small amplitude magnons in ferromagnetic spinor BECs via phase imprinting \cite{Marti2014}, and to selectively create spin waves without phase imprinting \cite{Kim2019}.

\begin{figure*} [htbp]
\includegraphics[width =2 \columnwidth]{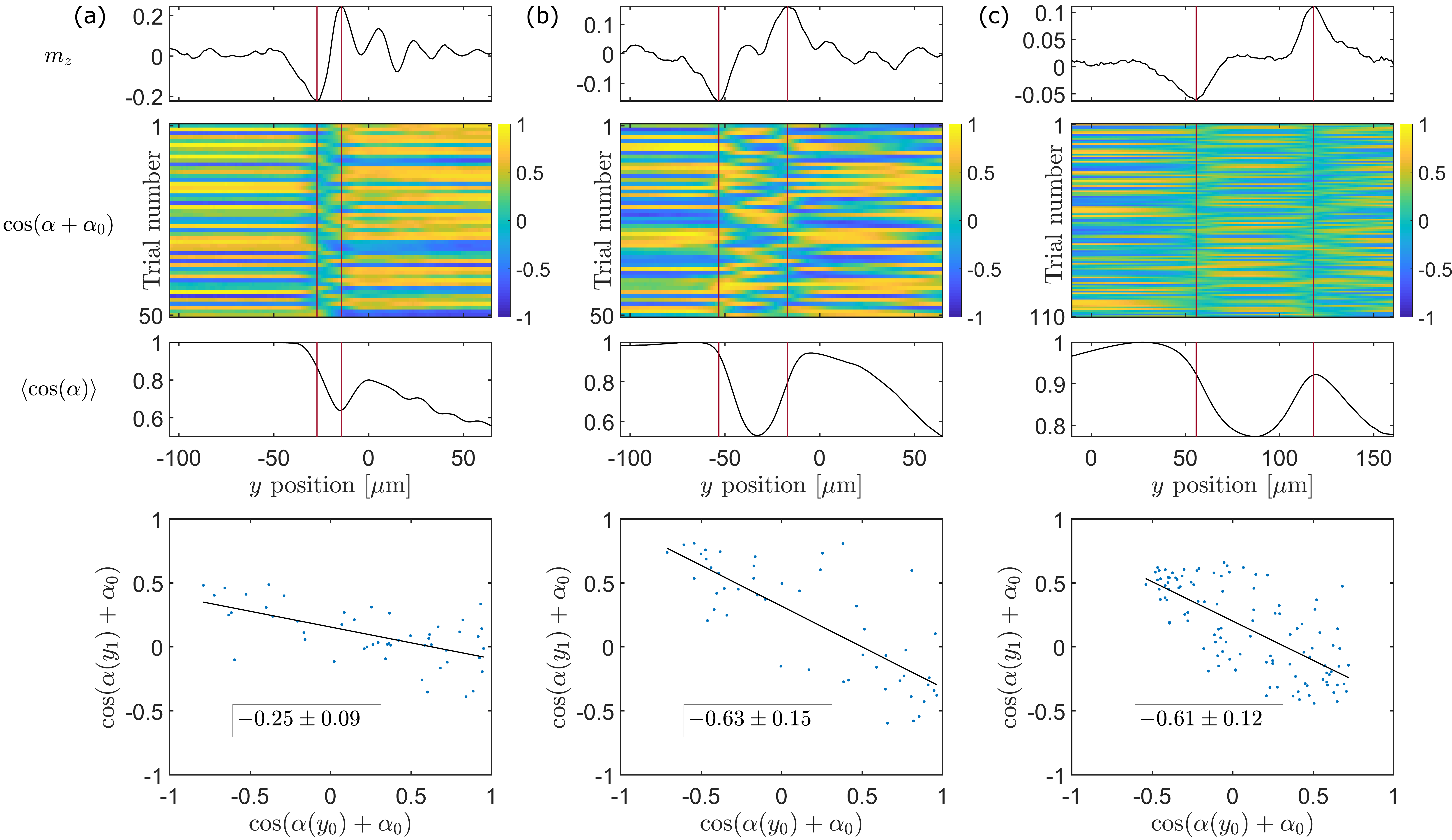}
\caption{(Color Online). The phase jump of the soliton.  (a), (b), and (c) Images (second row) show spatially resolved Ramsey spectroscopy (SRRS) measurements of $\cos({\alpha}(y)+\alpha_0)$, where $\alpha(y)$ is the phase difference between the two spin components and $\alpha_0$ is the background phase difference due to magnetic field fluctuation or field inhomogeneity, for (50, 50, and 110) independent runs and hold times of (0, 10, and 20) ms, respectively.  Uppermost plots show the position of the magnetic soliton pairs appearing in the corresponding magnetization data.  The third-row plots are $\langle \cos{\alpha(y)}\rangle$ with the background magnetic field noise subtracted (see Methods). The phase appears as a domain wall that divides the nematicity $N_{xx}$ into two regions whose separation grows with time.  Magnetic field noise prevented us from observing the full $\pm 1$ jump in $\cos{\alpha}(y)$ at each soliton boundary. The lowest plots show that the solitons are a domain wall across which the nematicity changes sign.  Plotted are the negative correlation coefficients for $\cos{\alpha}$ between points $y_0$ and $y_1$ to the left and right, respectively of the leftmost soliton. Each data point represents a single trial.  Here $y_0=(-67.34,-67.34,27.20)$ $\mu$m and $y_1=(-20.72,-34.97,86.77)$ $\mu$m, for (a), (b) and (c) , respectively.}
\label{fig:figure3}
\end{figure*}

Figure \ref{fig:main_result}(a) shows our principal data, where we have measured both the magnetization and phase profiles of the magnetic solitons.  In panel (a) we have plotted the time evolution of the one-dimensional magnetization $m_z(y,t)$ in the form of a spacetime diagram.  These data were normalized as $m_z(y,t) =1/2( n_1/n_1^0 - n_2/n_2^0)$ to the density profiles $n_i^0$ without phase imprinting to remove background density variations, including the Thomas-Fermi density profile of the cloud, as detailed in the Methods section.  It also rendered our data less sensitive to small differences from an equal spin mixture and to background magnetic field gradients.  The phase step becomes rapidly converted into a positive and a negative magnetization domain due to the equal and opposite momentum imparted to each spin state by the phase gradient.  Due to the very narrow transition region of $8 \mu$m, these magnetization domains have already formed within the finite duration phase imprinting pulse.  Once the pulse is over, these two domains, whose size is of the order of the width of the knife edge region, begin to separate from one another.  They propagate outward as a pair of solitons whose velocity $|V| = 1.20 \pm 0.12$ mm/s which is $\approx 0.82 c_s$ for our system.  Thus the solitons are seen to travel slower than the speed of spin sound.  According to Eqn.\ (\ref{eq:soliton}), the peak value of the magnetization pulse $|m_z| = \sqrt{1-U^2} \approx 0.26$ is consistent with our measured data.  Numerical simulations in 1 dimension are in very good agreement with our data. 

We also varied the pulse duration, and therefore the imprinted phase of the soliton, for 5 values--$70,120,170$ and 
$220 \mu$s at 0.5 mW laser power, and $80\mu$s at 0.8 mW laser power.  As the imprinted phase increased we observed the soliton magnetization to increase, while its velocity became slower.  The measured soliton velocity and magnetization shown in Figure \ref{fig:main_result}d show evidence of this weak inverse correlation, in spite of relatively large error bars.  The discrepancy between theory and experiment in Figure \ref{fig:main_result}(d) is likely due to the fact that time-of-flight absorption imaging reduced the measured magnetization contrast.



Our system is a two-component spinor embedded in an overall 3-component gas, and therefore has unique knobs for further probing of the soliton physics, particularly, its nontrivial phase profile.  Previous experiments have observed phase signatures in atoms outcoupled by Bragg scattering \cite{Denschlag2000}.  By contrast, we can employ atomic magnetometry in our system, a novel, {\em in-situ} probe to observe the soliton phase jump directly.  To this end, we used spatially resolved Ramsey spectroscopy (SRRS), as shown in Figure \ref{fig:figure3}.  This method utilizes the three-component nature of the spin vector to quantify the relative phase $\alpha = \alpha(y)$ between spin states $m_F = \pm 1$.  Just before time-of-flight, a second, fast RF $\pi/2$ pulse of 160 $\mu$s rotated each spin state about the $y$-axis of Figure \ref{fig:experimental_setup}(b).  We then performed a TOF-SG measurement that yields three spin populations $p_i(y) = n_i(y)/(n_1+n_0+n_{-1})$ for $i=1,2,3$.  An experimentally derived signal $S(y) = p_1+p_{-1}-p_0$, in effect, a measurement of the nematicity $N_{xx}$, directly determined $\cos{\alpha(y)} =  S(y)/\sqrt{1-m_z^2(y)}$, where $m_z(y)$ is the experimentally measured magnetization (see Methods).  The soliton's $\pi$ phase shift should result in a jump in $\cos{\alpha}$ between $\pm 1$ at each soliton location, a signature that the nematicity has been divided into two distinct regions.  

Figure \ref{fig:figure3} shows the result of this experiment for hold times of $0$, $10$ and $20$ ms in panels (a), (b) and (c), respectively.  The second-row panel contains, respectively, 50, 50 and 110 separate realizations of the experiment.  These were taken to average out bias magnetic field fluctuations that lead to variations in the absolute phase difference $\alpha_0$ between the two magnetically sensitive states.  In spite of these fluctuations, there is a clear discontinuity in the measured value of $\cos{\alpha}$ at the location of each soliton.  The spacing between solitons was 13, 36 and 62 $\mu$m for $T = 0,10$ and $20$ ms, respectively, as determined separately by magnetization measurements of the type shown in Figure \ref{fig:main_result} and plotted in the uppermost panels of the Figure.  The fourth-row panel demonstrates the negative correlation between measurements of $\cos{\alpha(y)}$ for two points $y_0$ and $y_1$ on opposite sides of the leftmost soliton (values for $y_0$ and $y_1$ are given in the Figure caption).  This negative correlation is consistent with a division of the nematicity $N_{xx}$ into two regions, as expected for magnetic soliton solutions \cite{Qu2016,Fujimoto2019}.  The measured negative slopes were greater than $-1$ due to residual magnetic field fluctuations and the finite contrast of the measurement of $S(y)$.  The third-row panel further confirms the division of the nematicity.  It shows an average of all the data with $\alpha_0$ separately measured and removed (see Methods).  It drops from 1 at the location of the left soliton to a minimum value between $0.6-0.8$, and rises back up at the location of the second soliton.  This minimum value could not reach $-1$ due to residual magnetic field fluctuations caused by spatially inhomogeneous fields of higher order than the gradient.  These data nonetheless show that the region between the two solitons has a nematicity different from the outside, and that the size of this region grows with time.




\begin{figure*} [htbp]
\includegraphics[width =2\columnwidth]{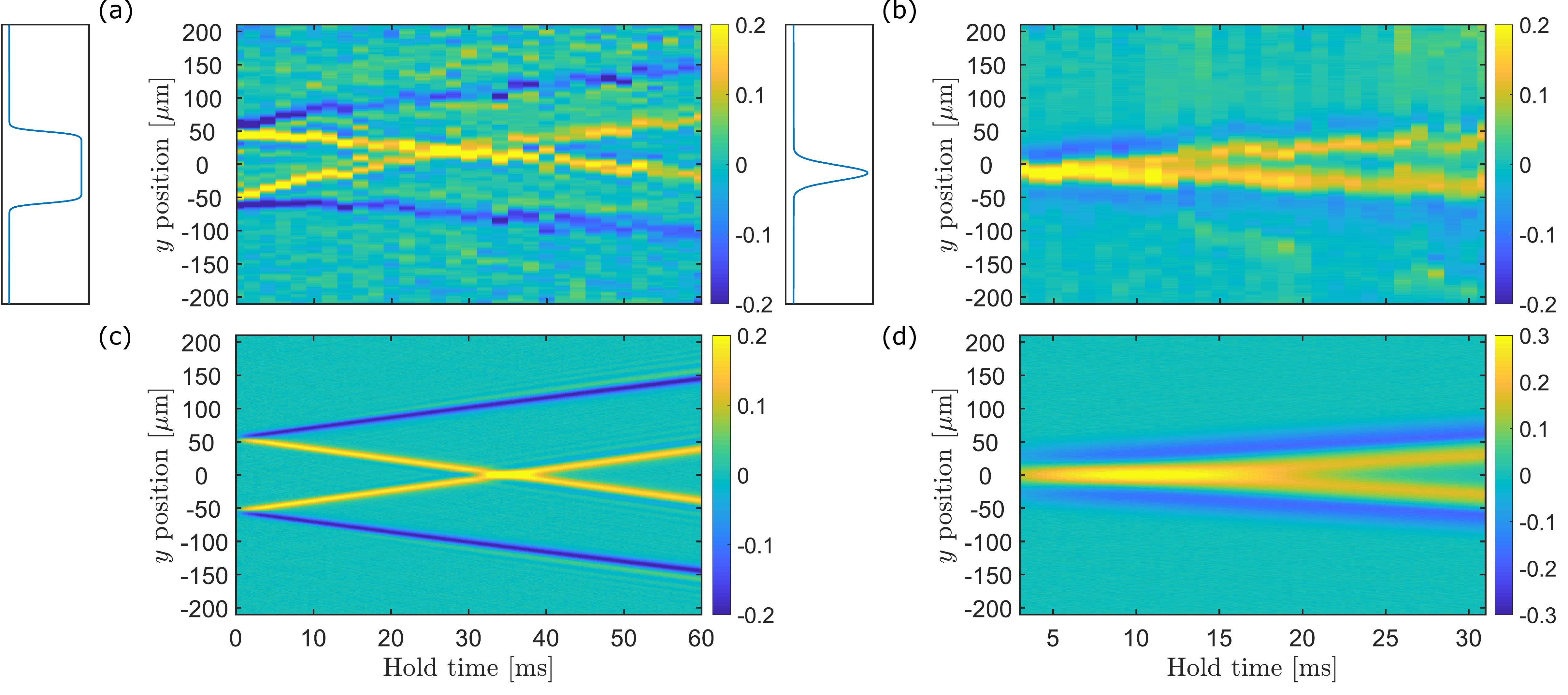}
\caption{(Color Online).  Engineering multiple soliton collisions at different spacetime points.  (a) Two knife edges created a flat-top beam with 110 $\mu$m width and 8 $\mu$m edges on both sides.  The magnetic shadow technique created two soliton pairs with mirror symmetric magnetization patterns.  Two positively magnetized solitons (yellow lines in the spacetime plot) passed through one another 25 ms after imprinting. (b) A tightly focused phase imprinting beam with 20 $\mu$m Gaussian waist created two soliton pairs starting from nearly the same location, so that the collision of the inner two positive solitons happens nearly immediately. (c) and (d) are the corresponding numerical simulations of the 1D GPE for the experimental parameters in (a) and (b), respectively. The phase profiles shown to the left of the data are fits to the measured laser beam profiles using a pair of spatially separated hyperbolic tangent functions (a) and a Gaussian beam (b).}
\label{fig:figure4}
\end{figure*}


A key advantage of the magnetic shadow technique is that one can engineer coherent magnetic structures with minimal disturbance to the condensate density profile.  Thus it is possible to create multiple magnetic solitons and to observe their interactions.  While dark solitons are generally expected to pass through one another undisturbed \cite{stellmer2008}, the behavior of magnetic solitons is not as well studied.  For these solitons, since the magnetization is independent from the phase jump, there are more possibilities for interactions, e.g., positive-positive vs negative-positive solitons.  The latter have been predicted to form bound states that can result in annihilation of the pair \cite{Fujimoto2019}.  

We demonstrate this capability in our experiment by placing 4 solitons in close proximity near the center of the condensate, and observing the interaction between solitons of the same sign of magnetization, and opposite velocities.  Figure \ref{fig:figure4} shows the experimental result and numerical simulation for two types of soliton collisions occurring at different times.  In the first experiment, two knife edges were used to create a flat-top beam of 110 $\mu$m width and 8 $\mu$m edges.  Due to the mirror symmetry of the configuration, two positively (negatively) magnetized solitons propagated inward (outward).  The inward propagating solitons then encountered each other at a time of 25 ms, passing through undisturbed.   In the second experiment a tightly focused phase imprinting beam with 20 $\mu$m Gaussian waist created two soliton pairs starting from nearly the same location, so that the collision of the inner two positive solitons happens nearly immediately.  The resulting magnetization for longer times resembled two pairs of co-propagating magnetic solitons, where each consists of a positive and negative magnetization peak. 
In all cases, the 1D simulations showed very good agreement with the measured results.

\section{Discussion}

Magnetic solitons in a highly elongated spin-1 BEC were created by the method of magnetic phase imprinting, and good agreement was observed with numerical simulations based on the 1-dimensional Gross-Pitaevskii equation.  Unlike turbulent methods of creating solitons, our technique of ``magnetic shadowing'' allows for coherent spin structures to be created and followed dynamically with very high purity (no spurious phonon creation, for example).  It opens up a host of novel studies in soliton physics not possible previously, including deterministic and highly controlled creation of solitons in large numbers (up to 100s) that comprise ``solitonic matter''.  This could be accomplished using multiple laser beams or, for example, optical lattices.  The creation of oppositely magnetized soliton pair collisions will enable the study of universal relaxation dynamics of spin-1 Bose-Einstein condensates \cite{Fujimoto2019}.  Our results open up a vista for a hitherto unrealized, fully quantum regime of solitons analogous to the quantum Hall effect for rotating bosons that has thus far remained elusive due to the centrifugal instability \cite{schw04}.  For example, addition of a two-dimensional lattice could allow us to reach the regime where the number of solitons and particles become comparable to one another, resulting in a strong correlation between atoms and a ``soliton quantum" of charge corresponding to the integrated magnetization.  These possibilities are extremely exciting as they can reveal the true quantum nature of the solitons.

\section{Methods}

Spinor condensates of $2.1 \times 10^7$ sodium atoms were created in the $m_F = 0$ state in a procedure documented in earlier work \cite{vinit2013,vinit2017}.  The cigar-shaped cloud with 70:1 aspect ratio was optically trapped with oscillation frequencies of $(\omega_x,\omega_y,\omega_z) = 2\pi \times (380,5.4,380)$ Hz, with the soliton dynamics occurring along the slow $y$-direction. The Thomas-Fermi radii are $(7.1,7.1,480)~\mu\mathrm{m}$. We refer to these as real space coordinates in contrast to the spin-space coordinates of Figure \ref{fig:experimental_setup}.  We prepared the atoms in a nominally equal superposition of spin states $m_F = \pm 1$ with residual $m_F = 0$ population $< 5\%$.  We did so by applying an RF pulse of 160 $\mu$s duration to the initial $m_F = 0$ condensate at $t = 0$.  Immediately after the RF pulse, we applied the magnetic shadow pulse.  The system then evolved for an additional time $T$ during which the solitons developed and separated, after which we measured the density profiles in each spin component using a 9 ms of time-of-flight Stern-Gerlach imaging sequence.  During the time evolution $T$, a far-detuned microwave field was kept on that stabilized the two component mixture, as described below.   We detail these protocols in the section below.

\subsection{Spin-state preparation and magnetic bias field setting}

For the three component spinor, the initial RF pulse can be viewed as a 90 degree rotation of the nematic director from the $z$ to $x$ axes about the $y$-axis using the spin-space coordinate system of Figure \ref{fig:experimental_setup}(b).  
In the absence of phase imprinting, the dynamics of this director are simply Larmor precession about the $z$-axis at a rate $\omega_L = g_F \mu_B B_0/\hbar$, where $B_0$ is the external bias field.  $B_0$ was determined by magnetic resonance imaging of the three spin components, and set using external Helmholtz coils to $70 \pm 4$ mG, resulting in a Larmor precession frequency of $\omega_L/2\pi = 50 \pm 3$ kHz.  Since our experiment lasted for up to 60 ms, a stability of $\approx 10~\mu$G would be required to ensure a constant phase across the cloud.  In practice, magnetic field gradients could only be cancelled to within 4.7 mG/cm, which resulted in a substantial phase winding for hold times $> 30~\mathrm{ms}$.  Therefore, for the phase sensitive data, we restricted our observations to a  small region 200 $\mu$m wide near the location of the knife edge where the phase profile was more or less uniform.  

Due to the antiferromagnetic nature of the sodium BEC, the quadratic Zeeman shift plays an important role in determining the stability of the two component mixture, and a first-order phase boundary at $q = 0$ separates polar ($q>0$) and antiferromagnetic ($q<0$) ground states \cite{Stamper-Kurn2013}.  Therefore, we applied a quadratic shift of $-35$ Hz using a microwave magnetic field detuned by $170$ kHz to the red from the $|F,m_F\rangle = |1,0\rangle$ and $|F,m_F\rangle = |2,0\rangle$ hyperfine clock transition at 1772 MHz, as in earlier work \cite{bookjans2011}.  The negative value of $q$ favored the two-component $m_F = \pm 1$ mixture, and prevented the $m_F = 0$ component from being substantially populated for the 60 msec duration of the experiment.  

As in our previous work, the tight radial confinement enabled largely one-dimensional spin dynamics of the soliton.  Nonetheless, the bias field $\vec{B}_0$ is a vector with three nonzero components.  Although in theory no preferred orientation for $\vec{B}_0$ exists absent dipolar interactions \cite{Stamper-Kurn2013}, in practice the influence of magnetic field gradients is not the same in all directions.  In particular, gradients along the long $y$-axis of the cloud play a more significant role than other directions since the spin dynamics along $x$ and $z$ are largely frozen.  For this work, we used a magnetic field $\vec{B}_0 = \hat{z}B_0$ perpendicular to the long axis of the cloud to enable the magnetic shadow imaging technique.  With this geometry, we could cancel background field variations along the important $y$-direction to within 0.3 mG.  Similar to earlier work, we held an $m_F = 0$ condensate for a period of 2 seconds in a uniform bias field with $q > 0$ and varied an applied field gradient $\partial B_y/\partial y$ to minimize the phase separation into $m_F=\pm 1$ \cite{vinit2017}.  Higher order field variations limited our observation time to 20 ms, as in Figure \ref{fig:figure3}. 

\subsection{Magnetic shadow}

The magnetic shadow shown in Figure \ref{fig:experimental_setup} was created by placing an ordinary razor blade in a far-off resonance laser beam, and imaging the shadow onto the center of the BEC.  The steepness of the intensity contrast (10\%-90\% width) was 8 $\mu$m at its focus, as measured by a parallel light path in free space utilizing the same lens configuration.
The laser beam was derived from a tunable dye laser operating between the sodium $D_1$ and $D_2$ lines whose splitting is $516$ GHz \cite{Steck}.  We set the laser frequency to $508.505$ THz, about $172$ GHz blue-detuned from the sodium $D_1$ transition, close to the magic wavelength where the scalar polarizability vanishes due to the opposite sign of $D_1$ and $D_2$ contributions \cite{Arora2007}.  At this wavelength, for circularly polarized light, the vector polarizability contributions of the two states have the same sign and interfere constructively, resulting in an effective magnetic field of $10$ mG for $2.5$ W/cm$^2$ laser intensity.  For our work we used a laser power of $0.5$ mW and a 120 $\mu$s pulse duration to realize a gradient of the magnetic field of $12$ mG and a differential Larmor phase shift of $2 \pi$ between the laser beam and the shadowed region.  The laser frequency drifted by no more than $\approx 2$ GHz over the course of the experiment, which had no discernible influence on the data.  We also checked that linear polarization, with no vector light shift, produced no magnetization in the cloud, as expected.

\subsection{Stern-Gerlach imaging}

In our experiment, all the data were collected by applying time-of-flight Stern-Gerlach imaging (TOF-SG) as seen in Figure \ref{fig:experimental_setup}. In order to spatially resolve the magnetic soliton, our imaging system consisted of a microscope objective with $10\times$ magnification and the calibrated resolution equal to 6.2 $\mu$m using a test target. During TOF-SG, the field gradient produced by a pair of anti-Helmholtz coils is turned on as well as and a bias field along the $x$-axis (perpendicular to the weak-axis of the dipole trap and gravity). This caused the cloud to separate principally along the $x$-axis.  After 9 ms, the three different spin components of the condensates separated completely, and a 25 $\mu$sec absorption imaging pulse was applied.  The duration of this pulse was chosen to maximize the imaging signal-to-noise ratio, while minimizing blurring due to atomic motion that reduced the measured magnetization contrast.  To overcome the difficulty with high optical density at short times-of-flight of only 9 ms, we carefully controlled the pump pulse detuning during imaging in order to keep the optical density below 1.  The net magnetization was obtained simply by subtracting the population of atoms in $m=\pm1$ states. However, spatially resolving the local magnetization of $m=\pm1$ spin components requires that the two condensates have to be perfectly aligned along the vertical direction in the images as shown in Figure \ref{fig:experimental_setup}(c) to prevent spurious magnetized structures from appearing.  We did so by using the mechanical effect of a  focused beam far red-detuned from the D1 transition to create a localized density stripe on {\em both} $m_F = \pm 1$ condensates, which provided us with a convenient and highly accurate way to align the two spin components. 

The data in Figures \ref{fig:main_result}-\ref{fig:figure4} were processed using the following protocol. Initially, we took two images for each run of experiment, one of which was illuminated by the phase imprinting beam and the other was not. Secondly, the data for each hold time were taken for 5 times in the same condition. Thirdly, the image with phase imprinting was normalized by the one without phase imprinting and then averaged with another four sets of data. Lastly, the spatially resolved magnetization for various hold times was extracted by subtracting the $m=+1$ condensate with $m=-1$ condensate and summing up the normalized magnetization profile of condensate along radial direction.  Center-of-mass oscillations of the BEC were removed by applying a sinusoidal fit to extracted center positions from individual bimodal fits to the expanded cloud.  These data are shown in Figures \ref{fig:main_result}-\ref{fig:figure4}.  The position $y_c$ and peak magnetization $m_z = \sqrt{1-U^2}$ of the solitons were extracted by fits to a magnetic soliton profile $\frac{n_0}{2}(1+\sqrt{1-U^2} {\rm sech} (\sqrt{1-U^2}(y-y_c)/\xi_{sp}))$ \cite{Qu2016}.  

\subsection{Spatially resolved Ramsey spectroscopy}

To show the evidence of the domain wall of two magnetic solitons, the technique called spatially resolved Ramsey spectroscopy (SRRS) was used in our experiment. The experimental sequence for this technique is the same as the measurement of the propagation of the solitons except for adding an extra $\pi/2$-pulse to rotate the nematic director another 90 degree just before the TOF-SG imaging sequence. In this way, the information of the relative phase between the $m=\pm1$ spin components can be decrypted from the populations of these three spin components. If we parametrize a generic easy-plane $F=1$ spinor \cite{Stamper-Kurn2013},
\begin{equation}
\psi(y)=
\begin{pmatrix}
\sqrt{\frac{1+m_z(y)}{2}}e^{i\frac{\alpha(y)}{2}} \\ 0 \\ \sqrt{\frac{1-m_z(y)}{2}}e^{-i\frac{\alpha(y)}{2}}
\end{pmatrix} \label{eq:spinor}
\end{equation}
where $m_z(y)$ is the magnetization and $\alpha(y)$ is the relative phase between $m=\pm1$ spin components. Then the final fractional populations $p_m(y) \equiv n_m(y)/(n_1(y)+n_0(y)+n_{-1}(y))$ after the second $\pi/2$-pulse can be readily shown to be 
\begin{eqnarray}
p_1(y)&= p_{-1} (y)=&\frac{1}{4}-\frac{1}{4}\sqrt{1-m_z^2(y)}\cos{\alpha(y)} \nonumber \\
p_0(y) & = & \frac{1}{2}+\frac{1}{2}\sqrt{1-m_z^2(y)}\cos{\alpha(y)} 
\label{eq:pops}
\end{eqnarray}
from which we define a composite experimental signal  $S(y) \equiv p_1(y)+p_{-1}(y)-p_0(y)$ and from  Eqs. (\ref{eq:pops}) above, 
\[ \cos{\alpha(y)} = \frac{S(y)}{\sqrt{1-m^2_z(y)}} \]
In terms of the nematicity ${N}_{ij}\equiv \frac{1}{2}\langle \hat{F}_i\hat{F}_j+\hat{F}_j\hat{F}_i \rangle$, it is easy to show from Eq.\ (\ref{eq:spinor}) that
\[N_{xx}(y) = \frac{1}{2}(1+\sqrt{1-m^2_z(y)}\cos \alpha(y)) = \frac{1}{2}(1+S(y)) \] 
We extracted the intrinsic phase jump across the magnetic soliton \cite{Qu2016}
\begin{equation}
\alpha(y) = {\rm Arctan} \left [ -\sinh{\left ( \zeta \sqrt{1-U^2} \right )}/U \right ]
\label{phase_jump}
\end{equation}
using the following procedure.  Here $\zeta = (y-y_c(t))/\xi_{sp}$ is the normalized soliton position in the moving frame.  Ambient magnetic field fluctuations contributed a random phase $\alpha_0$ from one run to the next, while higher order magnetic field variations were not possible to eliminate over the entire $960~\mu$m Thomas-Fermi diameter of the BEC.  Therefore, we restricted our observations to the 200 $\mu$m neighborhood of the solitons for hold times up to 20 ms.  We assumed that the phase to the left of the solitons has a constant value $\alpha_0$ due to the random phase fluctuation and the region inside the two solitons is $\alpha(y)+\alpha_0$, where $\alpha(y)$ contains all of the soliton phases induced by phase imprinting. Here we choose $\alpha_0$ at the locations $y=-67.34$ $\mu$m, $-67.34$ $\mu$m and $27.20~\mu\mathrm{m}$, relative to the origin at the cloud center in Figure \ref{fig:figure3}, for 0, 10ms and 20ms hold times, respectively. With $\cos( \alpha(y)) =\cos (\alpha(y)+ \alpha_0)\cos \alpha_0+\sin (\alpha(y) + \alpha_0)\sin \alpha_0$, we could eliminate the $\pm$ sign ambiguity of the sine function based on the assumption that the phase of the condensate varies continuously. Lastly, we averaged these processed phase data and obtain the results of $\langle \cos \alpha(y) \rangle$ as shown in Figure \ref{fig:figure3}.

\subsection{Numerical method}
We describe details of the numerical simulations for a system of $N$ spin-1 bosons. When the system undergoes Bose-Einstein condensation, it is described by the multicomponent order parameter $\psi_{m}(\bm{r},t)$ with the magnetic quantum number $m=1,0,-1$ which is governed by the spin-1 Gross-Pitaevskii equation:
\begin{eqnarray}
i\hbar \frac{\partial}{\partial t} \psi_{m} = \left( -\frac{\hbar^2}{2M} \nabla^2 + \frac{1}{2}M(\omega_x^2 x^2 + \omega_y^2 y^2 + \omega_z^2 z^2)  \right) \psi_{m} \nonumber\\ 
+ qm^2 \psi_{m} + g_{\rm d}n \psi_{m} + g_{\rm s} \sum_{n=-1}^{1} \bm{F} \cdot (\hat{\bm{F}})_{mn} \psi_{n}, \nonumber\\
\label{3D_SGP}
\end{eqnarray}
where $q$, $\omega_{\alpha} ~(\alpha=x,y,z)$, and $g_{\rm d}$ ($g_{\rm s}$) are a quadratic Zeeman energy, a trapping frequency, and the coupling constant of a density (spin) interaction.
The total density and spin density vector are defined by $n=\sum_{m=-1}^1 |\psi_m|^2$ and $\bm{F} = \sum_{m,n=-1}^1 \psi_m^* (\hat{\bm{F}})_{mn} \psi_n $ with the spin-1 matrix $\hat{\bm{F}}$.
In terms of the s-wave scattering lengths $a_0$ and $a_2$ for the total spin $F=0$ and $2$ of binary collisions, the coupling constants are given by $g_{\rm d}=4 \pi \hbar^2 (a_2 + 2a_0) /3M $ and $g_{\rm s}=4 \pi \hbar^2 (a_2 - a_0) /3M $.

The parameters corresponding to the experiment are $\omega_x= \omega_z = 2\pi \times 380~{\rm Hz}$, $\omega_y=2\pi \times 5.4~{\rm Hz}$, $q=- 2\pi \times 35~{\rm Hz}$, $a_0=47.36 a_{\rm B}$, $a_2=52.98 a_{\rm B}$, and $N=2.1\times 10^{7}$ with the Bohr radius $a_{\rm B}$. Employing the Thomas-Fermi (TF) approximation, we estimate the spin healing length $\xi_{\rm sp} \sim 0.87~{\rm \mu m}$, the TF radius $R_x = R_z = 7.0~{\rm \mu m}$ and $R_y=490~{\rm \mu m}$. A typical size of a magnetic soliton is about $10~{\rm \mu m}$, and thus the experimental system is effectively regarded as a one-dimensional system for magnetic solitons. Also, we obtain the chemical potential $\mu = 380~{\rm nK}$ at the center of the condensate. Then, integrating the local chemical potential over the $x$-$z$ plane under the TF approximation, the averaged chemical potential becomes $\mu_{\rm av} = \mu/2=190~{\rm nK}$.

We reduce the three-dimensional Gross-Pitaevskii equation \eqref{3D_SGP} to the one-dimensional one by assuming the following wavefunction:
\begin{eqnarray}
\psi_m (\bm{r},t) = \phi_m (y,t) F(x,z). 
\label{MWF1}
\end{eqnarray}
The function $F(x,z)$ has the TF density distribution in the tightly trapped directions, which is given by
\begin{eqnarray}
F(x,z) = 
\begin{cases}
    \displaystyle \sqrt{\frac{2}{\pi R_{\rm eff}^2} \left( 1 - \frac{x^2+z^2}{R_{\rm eff}^2}  \right)} & (x^2 + z^2 \leq R_{\rm eff}^2 ); \\
    \displaystyle    0  & ({\rm otherwise}), 
  \end{cases}
  \nonumber \\
\label{MWF2}
\end{eqnarray}
where $R_{\rm eff}$ is an effective radius of the condensate. Substituting Eqs.~\eqref{MWF1} and \eqref{MWF2} into Eq.~\eqref{3D_SGP} and performing intergration over the $x$-$z$ plane, we obtain the one-dimensional Gross-Pitaevskii equation:
\begin{eqnarray}
i\hbar \frac{\partial}{\partial t} \phi_{m} = \left( -\frac{\hbar^2}{2M} \frac{\partial^2}{\partial y^2} + \frac{1}{2}M \omega_y^2 y^2 \right) \phi_{m} + qm^2 \phi_{m} \nonumber \\ 
 + g_{\rm d}^{\rm (1d)} n^{\rm (1d)} \phi_{m} + g_{\rm s}^{\rm (1d)} \sum_{n=-1}^{1} \bm{F}^{\rm (1d)} \cdot (\hat{\bm{F}})_{mn} \phi_{n},
\label{1D_SGP}
\end{eqnarray}
where we define the one-dimensional total density and spin vector as $n^{\rm (1d)} = \sum_{m=-1}^1 |\phi_m|^2$ and ${\bm F}^{\rm (1d)} = \sum_{m,n=-1}^1 \phi_m^* (\hat{{\bm F}})_{mn} \phi_n$. The interaction coupling constants become $ g_{\rm d}^{\rm (1d)}  =  4 g_{\rm d}/3 \pi R_{\rm eff}^2$ and $ g_{\rm s}^{\rm (1d)}  = 4 g_{\rm s}/3 \pi R_{\rm eff}^2$. In this dimensional reduction, trivial constants have been ignored. As for the effective radius, we use $R_{\rm eff}=8.5~{\rm \mu m}$, which is determined by fitting the one-dimensional chemical potential $\mu_{\rm 1d}$ calculated by the TF approximation in Eq.~\eqref{1D_SGP} to $\mu_{\rm av}$.

We numerically solve Eq.~\eqref{1D_SGP} by using a quasi-spectral method, and calculate the time evolution of the spinor Bose gas starting from an initial state explained in the following. 
The experiment imprints the phase difference $g(y)$ between two wave functions $\phi_1$ and $\phi_{-1}$ to generate magnetic solitons. 
Then, we use the following initial wavefunction:
\begin{eqnarray}
\left(
    \begin{array}{c}
      \phi_1 \\
      \phi_0  \\
      \phi_{-1} 
    \end{array}
  \right)
  = \sqrt{ \frac{\phi_{\rm ini}}{2} }
  \left(
    \begin{array}{c}
     \displaystyle  (1+ \eta_{1} + i \eta_{2}) {e}^{ig(y)/2}\\
     \displaystyle   \eta_{3} + i \eta_{4}  \\
     \displaystyle  (1+ \eta_{5} + i \eta_{6}){e}^{-ig(y)/2} 
    \end{array}
  \right),
\label{initial_state}
\end{eqnarray}
where the noise $\eta_{j}~(j=1,2, \cdots, 6)$ is sampled by a Gaussian distribution with the mean $\mu=0$ and the standard deviation $\sigma=0.005$. 
The initial wave fucntion $\phi_{\rm ini}$ is obtained by the imaginary time-step method of Eq.~\eqref{1D_SGP} with the positive $q$. 
The phase difference $g(y)$ used in Fig.~\ref{fig:main_result}(b) is given by
\begin{eqnarray}
g(y) = \pi {\rm tanh}\left( \frac{y}{d_1} \right), 
\end{eqnarray}
where the width of the imprinted phase is $d_1=4.3~{\rm \mu m}$. 
In Fig.~\ref{fig:figure4}(a), we use the following function with the two phase jumps: 
\begin{eqnarray}
g(y) = 2 {\rm tanh}\left( \frac{l}{2d_1} - \frac{|y|}{d_1} \right),
\end{eqnarray}
where $l=110~{\rm \mu m}$. 
On the other hand, in Fig.~\ref{fig:figure4}(b), we use the Gaussian function defined by
\begin{eqnarray}
g(y) = 5 {\rm exp}\left( -2\left(\frac{y}{d_2} \right)^2 \right), 
\end{eqnarray}
where $d_2=23~{\rm \mu m}$. 

\section{References}


\section{Acknowledgments}

We acknowledge useful conversations with Carlos Sademelo and Colin Parker.  This work was supported by the National Science Foundation through award no. 1707654, and by KAKENHI  Grant No. JP18H01145, JP19K14628,  JP19H01824 and a Grant-in-Aid for Scientific Research on Innovative Areas ``Topological Materials Science" (KAKENHI Grant No. JP15H05855) from the Japan Society for the Promotion of Science. R. H. was supported by the Japan Society for the Promotion of Science through Program for Leading Graduate Schools (ALPS) and JSPS fellowship (JSPS KAKENHI Grant No. JP17J03189).

\section{Author contributions}
D. L. and X. C. contributed equally to data collection, discussions and experimental strategy.  D.L. played the major role in experimental setup and manuscript editing while X.C. played the major role in data analysis and figure preparation. C. Raman contributed to idea conception, data analysis, discussions and manuscript writing. K.F. contributed to idea conception and data analysis, and K.F., R.H. and M.U. contributed theoretically to numerical simulations, discussions and manuscript writing.

\section{Competing Interests} The authors declare no competing interests.

\end{document}